\documentclass[12pt,a4paper]{article}
\usepackage{epsfig}
\def\pslash{\rlap{\hspace{0.02cm}/}{p}}
\def\eslash{\rlap{\hspace{0.02cm}/}{e}}

\begin{document}
\title{Probing neutral top-pion via a flavor-changing process
$\gamma\gamma\rightarrow t\overline{c}\Pi_{t}^{0}$  }

\author{Xuelei Wang$^a$, Bingzhong Li$^b$,Yueling Yang$^a$  \\
  {\small a. College of Physics and Information Engineering,}\\
\small{Henan Normal University, Xinxiang  453002. P.R.China}
\thanks{This work is supported by the National Natural Science
Foundation of China(10175017), the Excellent Youth Foundation of
Henan Scientific Committee(02120000300), the Henan Innovation
Project for University Prominent Research Talents(2002kycx009).}
\thanks{E-mail:wangxuelei@263.net}\\
\small{b.Department of Physics, Tsinghua University, Beijing
100084, China}}
\maketitle
\begin{abstract}
\hspace{5mm}In the framework of topcolor-assisted-technicolor
model(TC2), we study a flavor-changing neutral top-pion production
process $\gamma\gamma\rightarrow t\overline{c}\Pi_{t}^{0}$. The
study shows that there exists a resonance effect which can enhance
the cross section up to a few fb even tens fb. For a yearly
luminosity 100 $fb^{-1}$ at future linear colliders, there might
be hundreds even thousands events to be produced. On the other
hand, the background of such flavor-changing process is very clean
due to the GIM mechanism in SM . With such sufficient events and
clean background, neutral toppion could be detected at future
linear colliders with high center of energy and luminosity. Our
study provides a possible way to test TC2 model.
\end {abstract}
\vspace{1.0cm} \noindent
 {\bf PACS number(s)}:12.60Nz, 14.80.Mz,.15.LK, 14.65.Ha
\newpage
\noindent{\bf I. Introduction}

 In the standard electroweak model, the $SU(2)_{L}\otimes U(1)_{Y}$local gauge symmetry is
 spontaneously broken to $U(1)_{em}$ through the medium of
 assistant,elementary scalar fields known as Higgs bosons. The
 self-interaction of the Higgs scalars select a vaccum, or minimum
 energy state, which does not hold the full gauge symmetry of
 the Lagrangian. With this, they endow the gauge bosons and the
 elementary fermions of the theory with masses. In the minimal
 model, three of the four assistant scalars become the longitudinal
 components of $W^{+}$,$W^{-}$ and $Z^{0}$. The fourth emerges as the physical Higgs
 boson, which has been the object of our attention in both theory and
 experiment. Though the phenomenological deduction of the standard
 model(SM) is very successful, the elementary scalar solution to
 the spontaneous symmetry breaking may be critized as arbitary, ambiguous ,or even
 theoretically inconsistant. There are so many
 free parameters associated with the Higgs potential and the
 Yukawa couplings that generate fermion masses. In addition, the  Higgs boson has not been
 found yet. Therefore, it is fair to say that the true mechanism of
 electroweak symmetry breaking(EWSB) in nature is still unknown.
 One hopes for a better, more restrictive solution, with greater
 predictive power.

 A new strong dynamics theory, technicolor(TC) model introduced by Weinberg\cite{y1} and
 Susskind\cite{y2}, offers a new insight into possible
 mechanisms of electroweak symmetry breaking. This kind of strong
 dynamical models of EWSB have evolved in the past dozens of
 years. As late as 1990s, we arrived at viable models in which a
 new dynamic, topcolor,  can coexist and the top quark acquirs a
 dynamical mass through topcolor. Such model, called topcolor assisted
 technicolor model(TC2)\cite{y3}, predicts a rich phenomenology that may be accessible to the colliders.

 In TC2 model, the EWSB is driven
mainly by technicolor interactions, the extended technicolor gives
contributions to all ordinary quark and lepton masses including a
very small portion of the top quark mass: $m^{'}_{t}=\varepsilon
m_{t}$ $(0.03\leq \varepsilon \leq 0.1)$\cite{y4}. The topcolor
interaction also makes small contributions to the EWSB and gives
rise to the main part of the top quark mass:
$(1-\varepsilon)m_{t}$ . One of the most general predictions of
TC2 model is the existence of three Pseudo-Goldstone Boson, so
called top-pions: $\Pi^{0}_{t}$,$\Pi^{\pm}_{t}$, which masses are
in the range of hundreds of GeV. The physical top-pions can be
regarded as the typical feature of TC2 model. Thus, the study of
the possible signature of top-pion and top-pion contributions to
some processes at the high energy colliders is a good manner to
test TC2 model.
 There have been many papers concerned with this subject
\cite{y5}. Another feather of the TC2 model is that topcolor
violates the GIM symmetry by treating the interaction of third
family differently from those of first and second. This
non-universal interaction results in a significant flavor mixing
of top and charm quarks and the neutral top-pion can couple to top
and charm when one writes the interactions in the quark mass
eigen-basis\cite{y6}. So, the study of some processes involving
$\Pi^{0}_{t}t\bar{c}$ vertex can provide information on
flavor-changing neutral current. On the other hand, such processes
can provide a good way to probe neutral top-pion at high energy
colliders. Recently, we have
   studied a flavor-changing neutral top-pion production process $e^{+}e^{-}\to
   t\bar{c}\Pi^{0}_{t}$\cite{y7}. We find that the resonance effect
   can enhance the cross section significantly when top-pion mass is
   small. Ref\cite{cao} has studied a top-charm associated production process at LHC
   to probe top-pion.  The above studies provide the feasible ways to detect
   top-pion events and test TC2 model.

The future $e^+e^-$ colliders can also operate in the $e\gamma$ or
 $\gamma\gamma$ modes. High energy photons for $\gamma \gamma, e\gamma$ collisions
can be obtained using compton backscattering of laser light off
the high energy electrons. In this case, the energy and luminosity
of the photon beam would be the same order of
 magnitude of the original electron beam and the set of final states
 at a photon collider is much richer than that  in the $e^{+}e^{-}$
 mode. Furthermore, one can vary polarizations of photon beams
 relatively easily, which is  advantageous for
 experiments. So, the future high energy and luminosity photons collides can provide us another way to test the new physics.
 In Ref.\cite{wang}, we studied the neutral top-pion
 production processes $e\gamma \rightarrow e\Pi_t^0$ and $\gamma \gamma \rightarrow
 t\bar{t}\Pi_t^0$ at photon colliders.
 The results show that, with favorable parameters, the cross section can reach the level of
 a few fb for the process $\gamma \gamma \rightarrow
 t\bar{t}\Pi_t^0$ and several tens fb for the process
 $e^{-}\gamma\rightarrow e^{-}\Pi^{0}_{t}$.
 $\Pi^0_t \rightarrow t\bar{c}$ is the most ideal channel to
detect neutral top-pion due to the clean SM background. On the
other hand, we can easily distinguish the neutral top-pion from
other neutral Higgs bosons in SM and MSSM via the process
$e^{-}\gamma\rightarrow e^{-}\Pi^{0}_{t}$. In this paper, we study
another flavor-changing neutral top-pion
 production process $\gamma \gamma \rightarrow
 t\bar{c}\Pi_t^0$ at photon colliders. The study shows that the
 cross section of the process can reach the level of a few fb even
 tens fb. It is possible to detect the neutral top-pion at future high energy and
 luminosity linear colliders(such as: TESLA).

\noindent{\bf II The calculation of the production cross section}

As it is known, the couplings of top-pions to the three family
  fermions are non-universal and the top-pions have large Yukawa couplings
  to the third generation.
 The relevant flavor-changing vertices including the large
top-charm transition can be written as\cite{y6}
\begin{eqnarray*}
i\frac{m_{t}\tan\beta}{v_{\omega}}K^{tc}_{UR}K^{tt^{*}}_{UL}
\overline{t_{L}}c_{R}\Pi^{0}_{t}+h_{.}c_{.}
\end{eqnarray*}
 where $\tan\beta=\sqrt{(\frac{v_{\omega}}{v_{t}})^{2}-1}$,
$v_{\omega}=246$ GeV is the EWSB scale and $v_{t}\approx 60-100$
GeV\cite{y6} is the top-pion decay constant. $K^{tt}_{UL}$is the
matrix element of the unitary matrix $K_{UL}$ which the CKM matrix
can be derived as $V=K^{-1}_{UL}K_{DL}$ and $K^{ij}_{UR}$ are the
matrix elements of the right-handed rotation matrix $K_{UR}$.
Their values can be written as:
 $$K^{tt^{*}}_{UL}=1 \hspace{2cm}K^{tt}_{UR}=1-\varepsilon$$
$$K^{tc}_{UR}\leq\sqrt{1-(K^{tt}_{UR})^{2}}\approx
\sqrt{2\varepsilon-\varepsilon^{2}}$$
 With the
flavor-changing coupling $\Pi^{0}_{t}-t-\bar{c}$, the neutral
top-pion can be produced associated with a single top quark at
$\gamma\gamma$ colliders.  The Feynman diagrams of the process is
shown in Fig.1 in which the cross diagrams with the interchange of
the two incoming photons are not shown.
\begin{figure}[h]
\begin{center}
\begin{picture}(150,100)(0,0)
\put(-100,-230){\epsfxsize125 mm \epsfbox{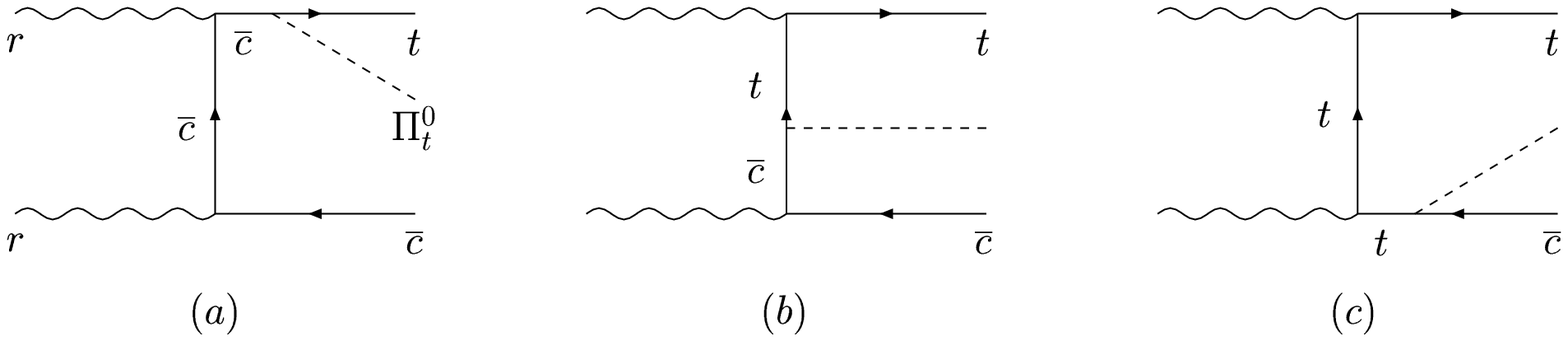}}
 \put(-80,-10){Fig.1 The Feynman diagrams for
 $t\overline{c}\Pi^{0}_{t}$ production in $\gamma\gamma$-collisions}
\end{picture}
\end{center}
\end{figure}\\
The amplitudes for the process  are given by
\begin{eqnarray*}
M^{(a)}&=&C\cdot G(p_{3}+p_{5},m_{c})G(p_{2}-p_{4},m_{c})\\
& &
\overline{u}_{t}(p_{3})R(\pslash_{3}+\pslash_{5}+m_{c})\eslash(e_{1})
(\pslash_{2}-\pslash_{4}+m_{c})\eslash(e_{2})v_{c}(p_{4})\\
 M^{(b)}&=&C\cdot G(p_{3}-p_{1},m_{t})G(p_{2}-p_{4},m_{c})\\
 & &
\overline{u}_{t}(p_{3})\eslash(e_{1})(\pslash_{3}-\pslash_{1}+m_{t})R
(\pslash_{2}-\pslash_{4}+m_{c})\eslash(e_{2})v_{c}(p_{4})\\
 M^{(c)}&=&-C\cdot G(p_{3}-p_{1},m_{t})G(p_{4}+p_{5},m_{t})\\
 & &
\overline{u}_{t}(p_{3})\eslash(e_{1})(\pslash_{3}-\pslash_{1}+m_{t})
\eslash(e_{2})(\pslash_{4}+\pslash_{5}-m_{t})R v_{c}(p_{4})
\end{eqnarray*}
The amplitudes of the diagrams with the interchange of the two
incoming photons can be directly obtained by interchanging
$p_1,p_2$ and $\eslash(e_{1}),\eslash(e_{2})$ in above amplitudes.\\
Where
\begin{eqnarray*}
C=-\frac{16\sqrt{2}}{9}\frac{m_{t}tan\beta}{v_{w}}M_{Z}^{2}
G_{F}c_{w}^{2}s_{w}^{2}K^{tc}_{UR}K^{tt^{*}}_{UL}
\end{eqnarray*}
and $G(p,m)=\frac{1}{p^{2}-m^{2}}$ is the propagator of the
particle, $L=\frac{1}{2}(1-\gamma_5)$,
$R=\frac{1}{2}(1+\gamma_5)$,
$s_{w}^{2}=sin^{2}\theta_{w},c_{w}^{2}=cos^{2}\theta_{w}$
($\theta_{w}$ is the Weinberg angle)

Using above amplitude, we can directly obtain the cross section
$\hat{\sigma}(\hat{s})$ for the subprocess
$\gamma\gamma\rightarrow t\overline{c}\Pi_{t}^{0}$, the total
cross section at the $e^+e^-$ linear collider can be obtained by
folding $\hat{\sigma}(\hat{s})$ with the photon distribution
function which is given in Ref\cite{y8}
\begin{eqnarray*}
\sigma_{tot}(s)=\int^{x_{max}}_{x_{min}}dx_{1}\int^{x_{max}}_{x_{min}
x_{max}/x_1}dx_{2} F(x_{1})F(x_{2})\hat{\sigma}(\hat{s})
\end{eqnarray*}
where $s$ is the c.m. energy squared for $e^+e^-$ and the
subprocess occurs effectively at $\hat{s}=x_1x_2s$, and $x_i$ are
the fraction of the electrons energies carried by the photons. The
explicit form of the photon distribution function $F(x)$ is
\begin{eqnarray*}
\displaystyle F(x)=\frac{1}{D(\xi)}\left[1-x+\frac{1}{1-x}
-\frac{4x}{\xi(1-x)}+\frac{4x^2}{\xi^2(1-x)^2}\right],
\end{eqnarray*}
with
\begin{eqnarray*}
\displaystyle
D(\xi)&=&\left(1-\frac{4}{\xi}-\frac{8}{\xi^2}\right)
\ln(1+\xi)+\frac{1}{2}+\frac{8}{\xi}-\frac{1}{2(1+\xi)^2},\\
x_{max}&=&\frac{\xi}{1+\xi}, \ \   \
\xi=\frac{4E_0\omega_0}{m^2_e}.
\end{eqnarray*}
where $E_0$ and $\omega_0$ are the incident electron and laser
light energies. To avoid unwanted $e^+e^-$ pair production from
the collision between the incident and back-scattered photons, we
should not choose too large $\omega_0$. This constrains the
maximum value for $\xi=2(1+\sqrt{2})$. We obtain
\begin{eqnarray*}
x_{max}=0.83, \  \   \      \  D(\xi)=1.8
\end{eqnarray*}
The minimum value for x is then determined by the production
threshold,
\begin{eqnarray*}
x_{min}=\frac{\hat{s}_{min}}{x_{max}s}, \ \  \  \
\hat{s}_{min}=(m_t+m_c+M_{\Pi})^2
\end{eqnarray*}
 \noindent{\bf III The results and conclusions}

In the calculations, we take $m_{t}=174$ GeV, $m_{c}=1.5$
GeV,$v_{t}=60$ Gev, $M_{Z}=91.187$ GeV, $s^2_w=0.23$. For
$K^{tc}_{UR}$, we take its maximum value, i.e., $K^{tc}_{UR}=
\sqrt{2\varepsilon-\varepsilon^{2}}$. There are three free
parameters involved in the production amplitudes, i.e.,
$\varepsilon,M_{\Pi} $ and the center-of-mass energy $\sqrt{s}$.
To find the effect of these parameters on the production cross
section, we take the mass of top-pion $M_{\Pi}$ to vary in certain
ranges 150 GeV$\leq M_{\Pi}\leq$ 350 GeV and
$\varepsilon=0.03,0.06,0.1$. Considering the planned $e^+e^-$
linear colliders (for example:
 TESLA), we should take $\sqrt{s}$=500 GeV, 800 GeV, 1600
 GeV . When $M_{\Pi}> 200$ GeV, $\sqrt{s}=$500 GeV
 is near to the production shreshold of $t\bar{c}\Pi^0_t$, the
 cross section is very small, we do not calculate the result for $\sqrt{s}$=500
 GeV. The results for $\sqrt{s}$=800,1600 GeV are summarized in
 Fig.2-4.

In Fig.2, taking $\sqrt{s}$=800 GeV  and
$\varepsilon=0.03,0.06,0.1$ respectively, we show the total cross
section of $t\overline{c}\Pi_{t}^{0}$ production as a function of
$M_{\Pi}$. We can see that the cross section increase sharply when
$M_{\Pi}$ become small, the reason is that there exists a
resonance when $M_{\Pi}<m_t-m_c$ which can enhance the cross
section significantly. For small $M_{\Pi}$, the cross section can
reach the level of several fb even tens fb.  With the yearly
integrated luminosity of $L\sim 100$ fb$^{-1}$ expected at future
linear colliders, hundreds even thousands events can be produced
via the process $\gamma\gamma \rightarrow t\bar{c}\Pi_t^0$. Since
the neutral top-pion with mass less than $2m_t$ decay mainly into
$t\bar{c}$ or $\bar{t}c$, the final state should be
$t\bar{t}c\bar{c}$ or $tt\bar{c}\bar{c}$. The final state
$t\bar{t}c\bar{c}$ is not a flavor changing in SM and may be
produced so much in SM. Such background analysis is difficult,
which is out of scope in our paper. But the final state
$tt\bar{c}\bar{c}$ is a flavor changing in SM, and the background
of $\gamma\gamma \rightarrow t\bar{c}\Pi_t^0\rightarrow
tt\bar{c}\bar{c}$ should be very clean. So, it is a better way to
probe the neutral top-pion via $tt\bar{c}\bar{c}$ signals. With
large cross section and clean background of signal
$tt\bar{c}\bar{c}$, the neutral top-pion should be detected at
future linear colliders, specially, in case of light
$\Pi^{0}_{t}$. In Fig.2, it is also shown that the cross section
increases with $\varepsilon$. To see the influence of $\sqrt{s}$
on the cross section, we give the numerical results for
$\sqrt{s}=1600$ GeV in Fig.3. The results show that the high
center-of-mass energy can enhance the cross section.

Due to the resonance effect for light toppion, there should be
   a peak near $m_t-m_c$ in the toppion-charm invariant
   mass distribution. To show the above results, we plot, in
   Fig.4, the toppion-charm invariant mass distribution with
   $\sqrt{s}$=800 GeV, $M_{\Pi}=$160 GeV, $\varepsilon=0.03$. This
   resonance peak can possibly be observable.
   So, the study of the invariant mass distribution may provide
   another good way to detect signal of toppion.

Comparing numerical results of the process: $\gamma\gamma
\rightarrow t\bar{c}\Pi_t^0$ with  that of $e^+e^- \rightarrow
t\bar{c}\Pi_t^0$ studied in Ref.\cite{y7}, we find that the cross
section of $\gamma\gamma \rightarrow t\bar{c}\Pi_t^0$ is much
larger than that of $e^+e^- \rightarrow t\bar{c}\Pi_t^0$ for the
same parameters, so, photon collision can provide us a better way
to search for neutral top-pion.

  In summary, we study a neutral top-pion production process
  $\gamma\gamma \rightarrow t\bar{c}\Pi_t^0$. The study shows that cross section of such
  flavor-changing process can reach the level of a few fb even
  tens fb. For light $\Pi_t^0$, the $c-\Pi_t^0$ channel resonance
  effect in Fig.1(c) can enhance the cross section significantly. There is
   a peak near $m_t-m_c$ in the toppion-charm invariant
   mass distribution. With such large cross section and clean
   background, it is possible to detect the neutral top-pion
   experimentally at future photon collider.

\newpage
\begin{figure}[ht]
\begin{center}
\begin{picture}(250,200)(0,0)
\put(-80,-70){\epsfxsize 140 mm \epsfbox{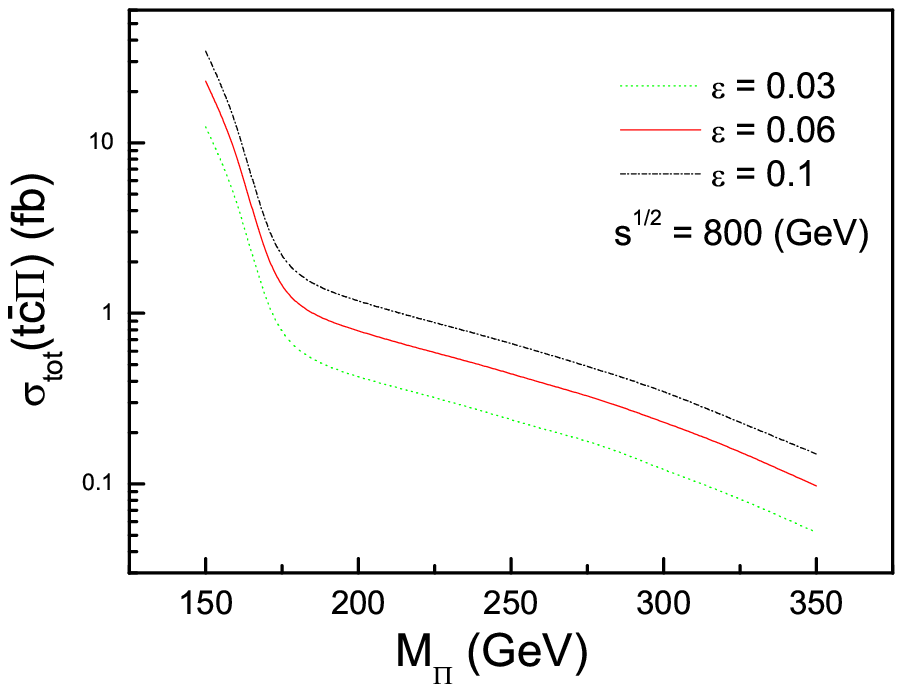}}
\put(-20,-70){Fig.2.  The total cross sections of
$t\bar{c}\Pi_t^0$ production versus the toppion }
\put(-40,-90){mass $M_{\Pi}$ for the center-mass-energy
$\sqrt{s}=800$ GeV
           and $\epsilon=0.03, 0.06, 0.1$.}
\end{picture}
\end{center}
\end{figure}

\begin{figure}[hb]
\begin{center}
\begin{picture}(250,200)(0,0)
\put(-80,-180){\epsfxsize 140 mm \epsfbox{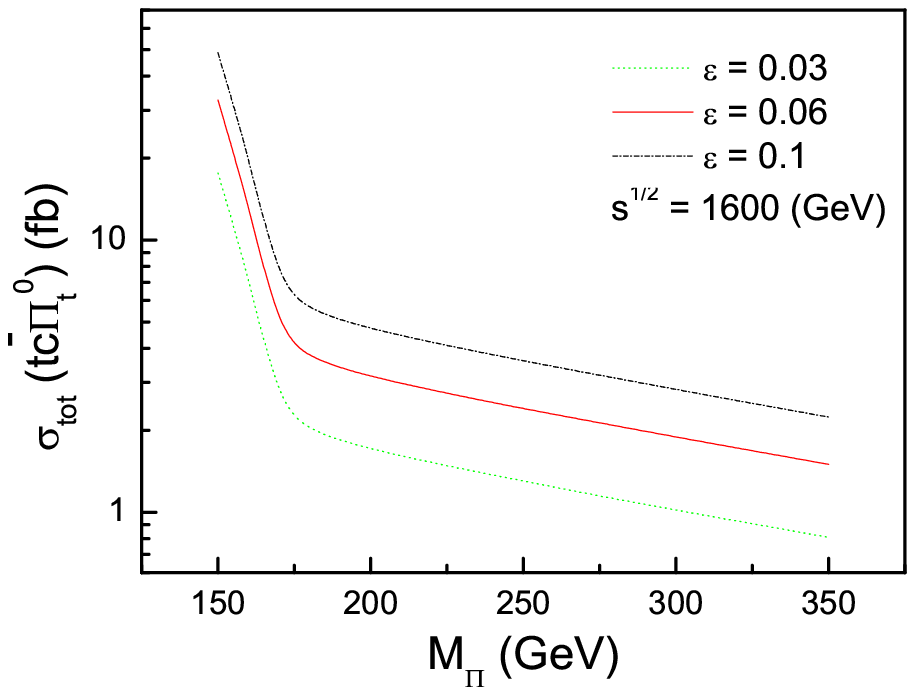}}
\put(-10,-180){Fig.3. The same plots as Fig.2 but for
$\sqrt{s}=1600$ GeV}
\end{picture}
\end{center}
\end{figure}

\newpage
\begin{figure}[ht]
\begin{center}
\begin{picture}(250,200)(0,0)
\put(-80,-70){\epsfxsize 140 mm \epsfbox{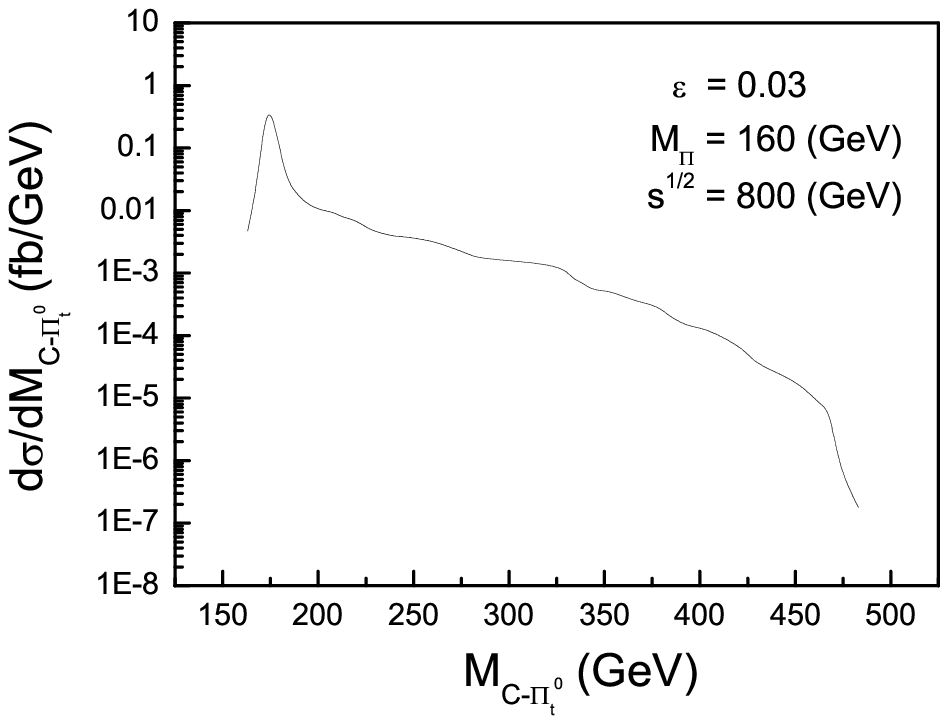}}
\put(-20,-70){Fig.4.  The topion-charm invariant mass distribution
for$\sqrt{s}=800 GeV$, } \put(-40,-90){ $\epsilon=0.03$ and
$M_{\Pi}=160 GeV$}
\end{picture}
\end{center}
\end{figure}

\end{document}